# Boundedness and Self-Organized Semantics: on the Best Survival Strategy in an Ever –Changing Environment

Maria K. Koleva

Institute of Catalysis, Bulgarian Academy of Sciences
1113 Sofia, Bulgaria

e-mail: mkoleva@bas.bg

*General criterion for best efficiency of the interaction of a complex system with an ever-changing environment is derived. Its exclusive property, set by boundedness, is that the highly non-trivial interplay between parameters that participate in it renders the best survival strategy to go via non-extensive hierarchical super-structuring of the semantic-like response.*

## Introduction

The greatest challenge to modern interdisciplinary science is constituted by the efforts for establishing general concept for satisfactory explanation of complex systems behavior. That is so because the behavior of complex systems reveals the same pattern of behavior for each system namely a ubiquitous coexistence of universal and specific properties. This coexistence persists regardless to the nature and individual properties of the system and is observed in a wide range of phenomena such as earthquakes, properties of DNA sequences, ant colonies, human languages, currency exchange rate sequences etc. For example, the heartbeat of all mammals displays the same pattern; yet we can unambiguously recognize whether a given pattern comes from a cat or a human; moreover, the patterns are individual, i.e. we are able to recognize which cat it is. Thus, the question arises: why the coexistence of specific and universal properties is necessary? Moreover, as a rule these systems operate at specific to each system yet ever-changing environment. In the same line of suggestions comes the next question: if such general reason exists, does it provide better survival strategy in an ever-changing environment? The importance of this question comes out from the fact that the well-known strategies "survival of the fittest" and "intelligent design" are appropriate only in a constant hostile environment because they are very vulnerable to small variations of the environment. Also, no extremal conditions such as energy/entropy extremization are appropriate because the generic property of complex systems is that they are open, i.e. each of them permanently exchanges matter/energy/information with its environment. Thus, in an ever-changing environment the energy/entropy are not invariants and in result they cannot be subjects of extremization.

The great importance of the question about the best survival strategy in an ever-changing environment is deeply connected with the issue about the contradiction between the stability of any complex system and widespread reading of the empirical data. To elucidate this point it should be stressed on the fact that complex systems exhibit remarkable stability in the sense that both their specific and universal properties remain the same over the time. At the same time the reading of the universal properties reveals a persistent inconsistency with this stability. In order to make myself clear, let me consider each of the universal properties separately. First in this line comes the so called $1/f$ noise phenomena which implies that the power spectrum of each time series that represents a record of the response of a complex system comprises a continuous band whose infrared end fits the shape $1/f$ where $f$ is the frequency. An exclusive property of that shape is that the fit does not depend on the length of the time series. Then, since the integrated power spectrum is a measure for the variance of the variability in the response, the logarithmic diversity introduced by the shape $1/f$ implies that the fluctuations gradually grow up which in turn would result in system`s breakdown. Further, this gradual enlargement of fluctuations should have originated in physical processes which would give rise to new extra lines in the power spectrum which, however, would compromise the empirically observed monotonicity of the continuous band in the power spectrum. Thus, this brief remark elucidates the severe controversy between the stability of complex systems and its violation coming from the up-to-date reading of the shape of the continuous band in the power spectrum.

The next universal property of the complex systems is the so called fractality of the response. It is characterized by the following: the response at different time (spatial) scales is self-similar and the scale is discernable only by comparison with an artificially introduced ruler. Then, each scale is characterized by power dependence between the actual shape measured by that ruler and the shortest line between its ends. Here, the matter about the controversy between the stability, i.e. time-translational invariance, and the power dependence arises as follows: what makes the spatial (temporal) scales to be so strictly "ordered" that this order persists in an ever-changing environment. Indeed, in an ever-changing environment all shapes should vary wildly and thus it is to be expected to violate the self-similarity of the "order" in the temporal and spatial scales. However, the data display remarkable insensitivity of the power exponents and self-similarity to changing details of the response.

The next contradiction between time-translation invariance (stability) and the universal properties of the complex systems is the so called "power law" distribution. It implies that the distribution of the events at a number of systems and phenomena follows power dependence. This means that, on the contrary to the normal distribution, the power law one does not select any average. The inconsistency with the time translational invariance appears as follows: the central idea behind the traditional approach to the power law distributions is that the lack of any average renders its most pronounced property to be the scale invariance. Alongside, the same distributions are assumed time translational invariant, i.e. their scale invariance is supposed independent from the beginning of the underlying processes that proceed in any given system. Yet, the concurrence of the both properties is biased by the conflict between the circumstances for their performance: while the scale invariance implies spanning of a power law over finite time/space intervals of unrestricted size, the time translation invariance requires the power law to be defined on infinitesimal time intervals only. Next in this line comes the following contradiction: a power law distribution implies that the rate of exchanged by the stochastic deviations energy/matter does not converge to a steady value but gradually increases with the length of time window. Then, it would turn out that the systems subject to power law distributions are unstable. The expected instability, however, is in a strange contradiction with the apparent empirical stability of the systems where it has been observed.

The inconsistency between the stability of complex systems and the above reading of their properties poses the question about the predictability of the complex systems behavior. Its importance is rendered not only by methodological purposes whether a given theory is better than any its rival one but by the fact that our major goal is to find out the best survival strategy in an ever-changing environment. In order to answer this question we must be able to predict the future behavior of any complex system by certain means. The fundamental non-triviality of this problem arises from the fact that no detailed prediction is ever possible since the variability in the response does not imply fine-tuning to the environment, i.e. it happens at each and every environment. Yet, at the same time the remarkable stability suggests that the future behavior is predictable in certain means. The problem becomes even more complicated because the traditional reductionist approach even worsens the situation. Indeed, according to it, each and every phenomenon in Nature is describable by a simple rule which does not change. However, the recent discovery that simulated dynamical systems exhibits strong sensitivity to initial conditions whenever the control parameters (where the environmental impacts enters in a specific way) are fine-tuned to certain values renders the future behavior of such system unpredictable by any means. The seemingly technical problem is not overcome by rival approaches such as self-organized criticality because the persistent $1/f$ noise appears at the expense of releasing the relation impact-response from any determinism. The latter implies that any small perturbation can trigger both a tiny response and an avalanche of gigantic response so that there is no rule which one happens when.

Another group of traditional approaches to the complex systems behavior are the probabilistic ones. Here I briefly discuss their common aspects which render unpredictability of the future behavior of the complex systems. The major one is that the modeling of the interaction system-environment follows the linear response theory which dictates a linear relationship between the impact and the response. This viewpoint, however, renders any response of a system silent and straightforwardly dictated by the environment. Further, the stability of a system follows the central physical postulate that the system behaves stably if and only if it is in equilibrium. The equilibrium is a single state which appears as a global attractor for almost all initial conditions. Then, any deviation from equilibrium is a fluctuation which spontaneously relaxes to it. Further, according to the fluctuation-dissipation theorem, any environmental impact is equivalent to a specific fluctuation so that they both are related linearly. Therefore, the question of predictability looses its sense since such system would loose its identity.

The brief outline of the major problems set by the attempts to explain the properties of the complex systems in the frame of the traditional approach explicitly reveals their fail. Moreover, it demonstrates that the dilemma is whether to persist following the traditional reductionist approach set by the traditional physics or one should adopt another point of view. I argue that we should abandon the traditional approach and create an alternative grounded on a new insight on the notion of a law. To elucidate our central point, let me remind that the traditional notion of a law implies a quantified relation between certain specific for the system variables so that the relation reproduces itself on repeating the phenomenon. Yet, what is tacitly presupposed is that the environment is also reproduced on reproducing the phenomenon. Put it is other words, the environment does not play an active role in this framework. That is why the enormous success of physics is to be associated with the study of closed, isolated systems and systems artificially put at constant environment. However, the present subject of interest, namely complex systems are open systems, i.e. they exchange matter, energy and information with the environment and this environment is an ever-changing one so that its detailed reproducibility is never likely.

In result, any research effort aimed to provide insights on studying complex systems faces the following fundamental dilemma: whether the weakness is a temporary difficulty of our current knowledge about complex systems or one should put forward an entirely novel concept for description of the complex systems behavior. I adopt the second alternative and recently proposed a novel concept which successfully resolves the conflict between time translational invariance and scale invariance of the behavior of complex systems [1]. The explanatory power of this new theory lies in providing instruments for explaining the ubiquitous coexistence of specific and universal properties.

I have proved that an exclusive property of that approach is that the response consists of two-parts: a specific resilient to environmental changes self-organized pattern, called homeostasis, and a universal part whose characteristics are insensitive to the statistics of the environmental impact. A central result of that approach is that the characteristics of the homeostatic pattern and the noise are additively decomposed in the power spectrum so that this decomposition obeys constant in the time accuracy. Put it in other words, in the frame of the proposed approach, one is able to reproduce any homeostatic pattern with the same accuracy even though the environment does not reoccur. It should be stressed that this result is generic for the concept of boundedness and does not require fine-tuning, i.e. it happens at all values of the control parameters. Thus, boundedness appears to be the first ever theory whose generic property to provide predictability of the future behavior by means of a successful reproduction of the characteristics of any homeostatic pattern.

Other central for the theory of boundedness result is that the ubiquitous coexistence of universal and specific properties opens the door to hierarchical semantic-like organization of the response. Here, semantic-like implies that the response follows internal rules so that to maintain the current homeostasis intact as long as possible. In result, the response does not follow silently the environment. Put it in other words, the response is not random even in a random environment. At the same time, the corresponding system is supposed adaptive, so that the response locally is able to "change patterns" but only in certain internally set order.

A very important branch of the developed approach, namely that of setting principles of hierarchical self-organization of complex systems driven by rules organized in semantic-like way, puts forward an alternative to the traditional information theory. Thus, it opens the door for looking for the best strategy for the next-generation performance strategy of a circuit designed according to the established rule. It is worth noting that such a semantic-like circuit is fundamentally different from a Turing machine: a Turing machine is able to execute any algorithm at the expense of non-autonomous decoding of the corresponding output; put it in other words, a Turing machine needs an external mind for comprehending the results of executing an algorithm. On the other hand, an exclusive property of a semantic circuit is that it executes only algorithms compatible with its functionality but at the expense of an autonomous comprehending of the output.

Outlining, the major goal of the present paper becomes to derive general criterion for the best survival strategy in ever-changing environment. As it will become evident, such criterion is straightforwardly related to the notion of best organized intelligence. Its exclusive property is that the latter does not implies size relations such as "biggest brain" or "fastest computer".

The paper is organized as follows: since the present paper is a further development of the theory of boundedness, a brief overview of the necessary results is presented in section 1. The major results about the establishing best survival strategy are presented in section 2.

## 1. Boundedness and Self-Organized Semantics – Brief Overview

The goal of this section is a brief presentation of the concept of boundedness and its development necessary for derivation of the best survival strategy in an ever-changing environment which task is considered in the section 2.

The concept of boundedness has been introduced a decade ago in the following paper [2]. The hypothesis of boundedness consists of 1) a mild assumption of boundedness on the local (spatial and temporal) accumulation of matter/energy at any level of matter organization and 2) boundedness of the rate of exchange of such an accumulation with the environment. Further, the response of a stable complex system is supposed specific and local. The latter implies that it is determined by local impact and the local state alone and can be defined locally by means of a specific rule. It should be stressed that the latter statement implies that the environmental impact acts not only non-linearly on the system but also in a non-homogeneous way. This suggestion implements the idea that the central "goal" of any response is to maintain the homeostasis intact as much as possible. It is worth noting that the homeostasis is self-organized pattern whose complex structure and functionality are not reducible to a single state such as the equilibrium in the traditional physics.

Another important suggestion derived from the above viewed goal is that complex systems tend to self-organize in a hierarchy of functional relations whose goal is to diversify the response so that every level to respond to certain stimuli only. An exclusive property of that hierarchy is that it is organized non-extensively and it operates in both directions: bottom-up and top-down. This constitute its major difference with the pyramidal hierarchy which operates only is one direction; thus the traditional reductionist approach assumes the lower-level phenomena such as cosmological ones, elementary particles etc, fundamental in the sense that they should determine in a unique way live, social phenomena etc.

The systematic development of the concept of boundedness gives a credible reconciliation of the contradiction between time-translational invariance (stability) of the complex systems and scale-invariance of their universal properties. A fundamental consequence of this setting is that it gives a novel understanding of the notion of information and intelligence. Next I briefly remind some important for the considerations in the next section results on this topics.

## 1.1 Time-Translational Invariance vs. Scale-Invariance

The following exclusive for the boundedness mathematical result [1] turns decisive for the reconciliation of the time-translational invariance and the scale invariance: it is proven that the boundedness alone is necessary and sufficient condition for: additive decomposition of the power spectrum of the series that represent any response to a specific discrete pattern which characterizes homeostasis and a universal continuous band which fits the shape $1/f^{\alpha(f)}$ where $\alpha(f)$ is a continuous monotonically increasing function whose infrared end is always $\alpha\left(\frac{1}{T}\right)=1$ ($T$ is the length of the time series). The so derived fit does not depend on the statistics of the time series or on its length. In turn, this immediately implies constant accuracy for the re-occurrence of the homeostatic pattern. A highly non-trivial result is that the fit $1/f^{\alpha(f)}$ provides finite variance of the deviations from the homeostatic pattern on the contrary to the shape $1/f$ which renders divergence of the variance on increasing the length of the time series. The proof is straightforwardly grounded on the assumption that the state space obeys Euclidian metrics. The non-triviality of the matter about the role of the state space metrics is to be found in [1] and will be discussed in the next subsection.

The next issue is to demonstrate how fractality appears in the setting of boundedness. It is already intuitively clear that one should associate fractalness with the deviations from the homeostatic pattern which appear as a result of the response to an ever-changing environment. Now there are two major interrelated issues: the first one what makes the "order of scales" and the second one what provides its robustness to permanent changes of the environment. Central for answering the first question is the exclusive property of bounded series that their coarse-grained structure is universal in the following sense: the coarse-grained structure of each and every bounded sequence comprises a succession of excursions of different size (up to thresholds of stability) so that each excursion is embedded in a finite interval specific to the size of an excursion so that no other excursion of the same or larger size is embedded in that interval. The "purpose" of embedding is that it maintains boundedness by means of preventing overlapping of successive excursions. Further, due to boundedness of the rates, the shape of each excursion is related to its "duration" by power law where the exponent is specific for the duration of any given excursion. It should be stressed on the fact that the duration of excursions is independent on the length of the time/space window of the observation; thus even the duration of the largest excursions is proportional to the thresholds of stability but not on the length of the window.

Now, it becomes evident that the "order of scales" appears due to the development of excursion of different sizes which, due to the embedding, are well separated from one another. Further, their property to be scale-free is revealed by the ubiquitous power dependence between the"duration", i.e. shortest line between the beginning and the end of an excursion, and the "shape" of the excursion. It is worth noting that namely this power dependence prevents selection of any specific time/space scale which in turn provides time-translational invariance of the response. The robustness of the above properties to the environmental changes is a result of the fact that the above described coarse-grained structure of bounded sequence is insensitive to the environmental statistics, i.e. it appears the same in an ever-changing environment.

Now another universal property of complex system, namely power law distributions will be discussed in the frame of boundedness. The concept of boundedness gives rise to two scenarios for their occurrence. The first one is to be associated with the distribution of the excursions, namely: the power law distribution appears as a heavy tail of the distribution of the excursions spanned over many orders of magnitude. The latter provides excellent fit by a specific to every given case power dependence. The second scenario appears as a result of using measures which are insensitive to the correlations introduced by higher order hierarchy. To make this statement clear let me consider the famous Zipf`s law which asserts: given some corpus of natural languages, the frequency of any word is inversely proportional to its rank in the frequency table. Thus the most frequent word will occur approximately twice often as the second most frequent word, three times as often as the third most frequent word, etc. Put it in other words, the Zipf law ignores any semantic meaning and thus seems to sweep out the difference between mind activity and random sequences of letters. It is a common knowledge that the semantics is permutation sensitive and puts a long-range order among its units. However, when considering the frequency of occurrence of a given word in a text, we ignore these higher level correlations and reduce the "system" to its simplest counterpart, i.e. to the first hierarchical level; and as we already have established, the trajectories in its state space is a dense transitive set of bounded sequences subject to power law distribution according to the first scenario.

## 1.2 Metrics in State Space and Hierarchy

One of the major premises of the concept of boundedness is that the stability is enhanced through diversification of the impact on a number of hierarchical levels. Then, a question arises: do the separation of the properties to specific and universal is a generic property of the hierarchical structuring. The non-triviality of the problem lies in the fact that a system must self-organize itself so that to meet boundedness of the transition rates to the locality of the response. This consideration is explicitly intertwined with the issue about robustness of the homeostasis: I suggest that it stays intact in a specific domain of control parameters space; passing into another domain it shifts to another pattern. The question is whether the property of partitioning the control parameter space into domains-of-attraction each of which is a specific pattern represented by an intra-basin invariant is matched onto the state space. I have proved that the answer to the above questions is affirmative and the separation of the response at each and every hierarchical level to specific and universal properties is a generic property for the hierarchical structuring.

The proof is grounded on the fact that boundeness of the rates renders one-to-one correspondence between the enumeration of the control parameter space and of the state space. Indeed, boundedness of the rates sets admissible only transitions to the nearest neighbors in the state space. Further, the distance to the nearest neighbors is set by the so-called inter-level feedbacks which appear as bounded stochastic term in the corresponding equation-of-motion. The equation-of-motion in the state space retains a generic property to be an effective reaction-diffusion bounded stochastic PDE of the following type:

$$\frac{dx}{dt} = \alpha A(x) - \beta R(x) + \nabla \bullet (D(x)\nabla x) + \mu(x) \tag{1}$$

where $\alpha$ and $\beta$ are the control parameters; $x$ is the "concentration" $A(x)$ is the "adsorption" function, $R(x)$ is the "reaction" function , $D(x)$ is the "diffusion" coefficient and $\mu(x)$ is the "noise" term which represents the inter-level feedback. It is worth noting that the "noise" term is subject to the mild constraint of boundedness alone; thus its distribution is left arbitrary.

In my book [1] I have proved that the power spectrum of the solution of eq. (1) comprises a part that comes from the solution of the "deterministic" part of eq. (1) along with a noise term. The "deterministic" part of eq. (1) reads:

$$\frac{dx}{dt} = \alpha A(x) - \beta R(x) + \nabla \bullet (D(x)\nabla x) \qquad (2)$$

It is well established that eq. (2) retains the generic property that its solutions partition the control parameter space into specific domains so that the solution in each domain appears as specific spatio-temporal pattern which is an intra-basin invariant. This justifies my suggestion to associate the homeostasis with those specific spatio-temporal patterns. Further, my suggestion that the homeostasis is not a single state is also justified by the fact that any given homeostatic pattern stays intact only in its domain, i.e. for a specific set of control parameter space.

A decisive for the entire approach circumstance is that it allows a non-ambiguous way of setting units for all parameters in eq. (1). The necessity for the choice of units in a unique way is set by the fact that the partitioning of the state space depends on the numerical values of the control parameters that participate in eq.|(1). In turn, the unique choice of units renders unique enumeration of the state space and thus opens the door to the question whether it sets certain metrics in the state space and what its purpose is. Luckily, there is a unique way of determination for the units that participate in eq. (1). It is set by the interplay between "reaction" "diffusion" and the thresholds over the bounded noise which represents the inter-level feedbacks. In turn, this renders a one-to-one correspondence between the enumeration of the control parameters space and the state space.

An immediate consequence of the above one-to-one correspondence is that it renders possible to densely "cover" the state space by a single-scale Voronoi tessellation so that each cell comprises nearest neighbors. Let me remind that a necessary condition for existing of metrics is the possibility for single-scale Voronoi tessellation of the corresponding space; and when each and every Voronoi cell comprises the current nearest neighbors, the metrics is locally Euclidian. Taking into account that every path in the state space under boundedness is realized on a latticised subset (whose details vary from one sample lattice to another), each latticised subset is equivalent to a sample of Voronoi tessellation. Keeping in mind that each cell in every sample of this Voronoi tessellation comprises the current nearest neighbors, one concludes that indeed a state space under boundedness retains metrics and that this metrics is locally Euclidian. It is worth noting once again that the metric is locally Euclidian because the property that each Voronoi cell comprises the current nearest neighbors is exclusive property of the boundedness. Now it becomes obvious the "purpose" of sustaining metric locally Euclidian: it provides the distance between nearest neighbors to be bounded over the whole state space.

It should be stressed on the fact that the role of the metrics in the state space is Euclidian is fundamentally different from the traditional role of metrics. In physics, metrics enters functionality of a system following the general frame that there exists structural pattern such that the system is in equilibrium and any deviation of that equilibrium is related to the corresponding functional properties by means of a specific relation between energy, entropy or information and the corresponding metrical deviations from the equilibrium pattern. The greatest success of that approach is strongly grounded on the use of the tensor approach for describing the relation between functional and structural properties since the power of the tensor approach lies in the fact that it provides time-translational invariance of any given rule.

The fundamental difference of the concept of boundedness from the above scenario is that the under boundedness the functional properties cannot be derived from the structural ones by means of recursive relations. The reason for this is that the functional properties are represented though solutions of eq. (1) whose major generic property is that besides specific homeostatic pattern and noise it also comprises an irrational component which comes from their interplay. Then, neither pattern can be algorithmically reached from any other by any recursive means. This, along with the fact that in an ever-changing environment neither system is subject to extremization, renders necessity for a novel scenario for the relation between structural and functional properties. Further, it should be stressed on the fact, that the environmental impact enters eq. (1) through control parameters that participate in it. Since neither solution of eq. (1) is proportional to the control parameters, the effect of the environmental impact turns non-linear and non-homogeneous on the contrary to the tensor approach which renders relations linear and homogeneous.

Outlining, it is established that there exists self-consistency between boundedness and the metrics in the state space: on the one hand boundedness provides setting of Euclidian metrics in the state space and on other hand the Euclidian metrics sustains boundeness for all possible basins-of-attraction. Thus, the state space acquires the generic property that its metrics is always locally Euclidian.

*1.3 Self-Organized Semantics*

So far we have considered two major topics: (i) how reconciliation between time translational invariance (stability) of specific properties and scale invariance of the universal ones appears without fine-tuning, i.e. for every choice of control parameters; (ii) our next task has been to demonstrate that this reconciliation holds for each level of the specific for a system hierarchical organization. These results constitute the task of the present subsection to demonstrate that hierarchical coordination is governed by a general protocol which is expressed in a semantic-like way and operates by grammatical-like rules.

The first step on this road is to remind that the hierarchical coordination enters the equation-of-motion at each and every hierarchical level through the participation of inter-level feedbacks in eq. (1); since the values of the thresholds of stability of the bounded noise which represents the inter-level feedbacks enter the units that define the numerical values of the control parameters in the equations-of-motion, it becomes obvious that the partitioning of the state space into specific set of basins-of-attraction is controlled by other hierarchical levels.

Further, the structure of eq. (1) retains the generic property that the motion in the state space is bounded by means of the role of the inter-level feedbacks: the boundedness of noise terms that represent the inter-level feedbacks renders the admissible transitions to be only those to the nearest neighbors in every basin-of-attraction. This property along with the partitioning of the state space into basins-of-attraction renders the admissible transitions from any given basin to be only that to the nearest neighbors. Further, the boundedness of the state space renders that basins-of-attraction tangent to a common point called by us accumulation point; alongside it renders the motion in the state space to be orbital. A direct implication of the above facts is the association of each intra-basin invariant with a "letter" and the accumulation point with "space bar" so that each orbit is a "word" separated by others through the space bar. And since the admissible transitions between letters are allowed only for nearest basins-of-attraction, the response is never a random sequence of letters.

My next task is to demonstrate how grammatical-like rules appear from the need of maintaining boundedness at any given hierarchical level. For the sake of clarity I will consider how a sentence arises from a sequence of words. My considerations are derived from considering certain generic properties of eq. (1). Let me start with the fact that sustaining boundedness implies ban over extra-accumulation of energy/matter/information at every point in the state space. In order to fulfill this task it is obviously that generically the "output" of "reaction" in any given locality should serve as the "input" of the "reaction" that proceeds in its neighborhood and is connected to it by the "flow". Next let me remind that the motion in the state space is orbital. Then, the consistency of the "chains" of specific reactions implies association of each orbit with a specific non-mechanical machine. A central for my further considerations is the proof that each non-mechanical engine is equivalent to a Carnot engine. Next considerations explicitly exploit the generic for Carnot engines property to be sensitive to permutations, namely in one direction a Carnot engine performs as a pump and in the other it performs as refrigerator. This fact along with necessity that successive engines are coupled by means that the output of any one serves as input to the next immediately renders that the succession of "words" in a sentence to be subject to "grammatical"-like rules. An immediate result reads that the sequence of "words" in the response is again irreducible to a random one.

The association of a semantic unit (word, sentence, and paragraph) with a specific non-mechanical engine poses the question whether it is ever possible to construct such perfect engine so that to let possible detailed prediction of the future behavior. This goal has two aspects: (i) the first one is to establish whether the equivalence to a Carnot engine implies also that the efficiency of a non-mechanical engine is bounded by the efficiency of the corresponding Carnot engine; (ii) the second aspect is whether it is ever possible to "drain" the environment in order to achieve the goal. The first aspect is considered in Chapter 9 in [1] where it is proven that the equivalence between a Carnot engine and its non-mechanical counterpart implies that the efficiency of the non-mechanical machine never reaches that of its Carnot counterpart. It is worth noting that this implies that a non-mechanical engine produces useful work only at the expense of undergoing specific structural and functional changes. An immediate consequence of that fact is that it is impossible to "transform" "noise" into "semantics".

In conclusion, I have proved that prediction of the future behavior in an ever-changing environment is possible in the sense that one can predict the behavior of the homeostatic pattern with constant in the time accuracy. This gives rise to the highly important question whether it is possible to regulate the structural and functional characteristics of a system so that to achieve best long-term survival strategy of the response. Now I am ready with the preliminaries and can provide the affirmative answer to this question.

## 2. Efficiency of a Self-Organized Semantics

The goal of the present section is to derive general criterion for the efficiency of a non-mechanical engine with respect to its interaction with an ever-changing environment. It should be stressed that the proof about equivalence between a non-mechanical and a Carnot engine involves the sum of useful work, i.e. work spent on 'semantic" part of the response and the energy spend on the interaction of a system with its environment. My task now is to prove that the ratio between the useful work and the energy spent to the interaction with the environment, called hereafter noise for short, goes through maximum. Its exclusive property is that it relates local properties of a cycle with hierarchical structuring in a highly non-trivial manner.

Given the size of a cycle in $(entropy, temperature)$ plane of the state space to be $L$; then the work spent on semantic part of the response is proportional to $L^2$. My next task is to work out the energy spent on the interaction of the corresponding system with its environment. Since the environment is supposed to be ever-changing and subject to the constraint of boundedness alone, it is measured by the length of the irregular bounded sequence "wrapped" onto $L$. Let us denote its thresholds of stability by $\Delta$. The derivation of this length is grounded on the following facts:
- Bounded irregular sequences have universal properties;
- One of that universal properties is that its coarse-grained structure is insensitive to the details of the statistics of the environment;
- Other universal property is that the Eulidianity of the metrics in the state space renders this structure neither to select nor to signal out any specific point. Put it in other words, the process of formation of excursions is a homogeneous one.

Then, it is to be expected that the length of the excursions has fractal properties whose major characteristics is their dimension. To remind, the latter serves as an exponent in a power dependence that relates the length of a fractal and the shortest line between its end points. In our case, it implies that the minimal length of "fractal" curve that represents the interaction of a system with its environment is approximated by the length of largest size excursions embedded in the working cycle. The lower bound of that length is given by $\Delta L$. This length comes out from presentation of each excursion as 1`st order discontinuity at every point. This presentation implies "smoothing out" any finer structure. It should be stressed that this approximation explicitly exploits the fact that the durations of excursions are independent from the length of the window thus immediately rendering the number of excursions to be proportional to $L$. Then, the Euclidianity of the metrics renders the 1`st kind discontinuities in every point to be also proportional to $L$. Then, the lower bound of length of the curve that represents the work spent of interaction with the environment is $\Delta L$. It should be stressed that the above listed properties of the bounded sequences render the length of self-similar pieces to be $\Delta L$, while the length of magnification is $L$. Then, the Hausdorff dimension of the fractal curve is:

$$D = \frac{\log(\Delta L)}{\log(L)} = 1 + \frac{\log \Delta}{\log L} \tag{3}$$

It should be stressed that $D$ is always greater than the topological dimension of the curve (which is 1) since both $\Delta$ and $L$, expressed in the corresponding units, must always be greater than the corresponding unit. This is so because: (i) $\Delta$ must be greater than the corresponding unit for the reason of providing free execution of "U-turns" at the thresholds. (ii) $L$ must be greater than 1 for the obvious reason that it must comprise changes of homeostasis which is implemented by a trajectory that goes via several basins of attraction. It is worth reminding that the units for each and every cycle are set in a unique way.

Thus, the efficiency of the semantic response $\eta$, expressed through the ratio between the work associated with changes in the homeostasis $L^2$ and the energy spent on the interaction with the environment $L^D$ reads:

$$\eta = \frac{L^2}{L^D} = \frac{L^2}{L^{1+\frac{\log \Delta}{\log L}}} = L^{1-\frac{\log \Delta}{\log L}} \tag{4}$$

Simple calculations show that the maximum efficiency of $\eta$ with respect to $L$ is set by the following relation:

$$\frac{\log \Delta}{\log L} = 0.25 \tag{5}$$

The ratio presented through eq. (5) quantifies the most efficient way of distributing the energy of a response: the best part of the energy goes to the "semantic" part of the response. In other words, the semantic response happens at the least possible expense of spending energy/matter in the interaction with the environment.

An immediate result of eq. (5) is that it gives rise to following generic approximation of the most efficient hierarchical organization:

$$\eta = L^{0.75} \tag{6}$$

One seems temped to conclude from eq. (6) that the best survival strategy is provided by means of a gradual increase of the size of working cycle $L$. Indeed, the sub-linear exponent $0.75$ renders the larger objects more advantageous because of their better efficiency of interaction with the environment.

However, this is an oversimplified interpretation of eq. (6). The non-triviality is revealed through the inter-dependence between $L$ and $\Delta$. This dependence appears as follows: any change in $L$ implies change in hierarchical structuring since the setting of basins-of-attraction at each level is controlled by neighboring levels through inter-level feedbacks. In turn, any change of $L$ is related to a change in $\Delta$. The exclusive property of that changes is they are not proportional one to another since the relations impact-response are non-linear and non-homogeneous. An immediate consequence is that the transformation of an object into a novel one happens in an allometric-like manner, i.e. parts of the same function that belong to different objects are not self-similar.

Another issue which demonstrates the high non-triviality of the intertwining between the size of a cycle $L$ and the thresholds of stability for the interaction with the environment $\Delta$ is that any attempt to isolate the cycle from the environment, i.e. putting $\Delta = 0$, yields infinite efficiency $\eta \to \infty$ as provided by eq. (4). However, this immediately implies building of a universal perpetuum mobile whose generic property is that its characteristics do not depend on $L$. Put in other words, the latter implies that all objects in the Universe are perpetuum mobile. Thus, it becomes clear that the interaction with the environment is crucial for diversification of the objects in the Universe.

The wisdom of the above considerations is that the most efficient survival strategy, along with being subject to the general rule given by eq. (6), goes via non-extensive allometric-like hierarchical super-structuring. In turn, this provides each its subject with a specific advantage in surviving. The exclusive property of that advantage is that it is not related to its size alone but depends in a non-extensive way on its hierarchical structuring. An immediate consequence of this result is that the more complex its hierarchical structuring is, the more stable it is. Thus, to certain extend unexpected, the intelligence which is always expressed by means of semantics, turns out to be the most powerful implement for providing the best survival strategy in an ever-changing environment.

**Conclusions**

I have started the paper by posing the question why the ubiquitous coexistence between specific and universal properties of complex systems is necessary. Now I am ready to answer this question. Put the answer in a nutshell this coexistence is necessary for providing diversity of objects so that each and every of them has specific advantage in surviving. This is due to the highly non-trivial interplay between the parameters that participate in the criterion (eq. (6)) for providing most efficient survival strategy in an ever-changing environment. This interplay justifies allometric-like relations between structural and functional properties of the similar parts belonging to different kinds of species. In consequence, the allometricity serves as implement for providing diversification of the species with respect to their survival properties. In turn, the obtained diversity of the corresponding response serves as an implement for enhancing stability of the hierarchical structuring. In conclusion, this makes the concept of boundedness self-consistent since the central for all its development suggestion is that the stability of the complex systems is provided by means of tightening the response through its diversification onto a hierarchy of levels so that their connectivity gives rise to semantic-like manner of response.

The initial purpose for introduction of the concept of boundedness has been to explain the ubiquitous coexistence of universal and specific properties of complex systems. Now it becomes evident that the concept of boundedness not only successfully manages to fulfill this goal but its present development renders it with the power of creativity as well. Indeed, the established in the present paper general principles about best survival strategy can serve as guides for setting a next generation performance strategy for creating a family of circuits able to autonomous creation of information and its autonomous comprehending and communication. The relevance of this goal is justified by the conclusion drawn from this paper that the intelligence, viewed as the most efficient hierarchical structuring, is the leading factor in providing best survival strategy.